# Substrate temperature-dependent dielectric and ferroelectric properties of (100) – oriented lead-free $Na_{0.4}K_{0.1}Bi_{0.5}TiO_3$ thin films grown by pulsed laser deposition

Krishnarjun Banerjee[#], Adityanarayan H. Pandey[*#], Pravin Varade, Ajit R. Kulkarni, Abhijeet L. Sangle, N. Venkataramani[†]

Department of Metallurgical engineering and Materials Science, Indian Institute of Technology Bombay, Mumbai-400076, India.

**Abstract:** Pb-free ferroelectric thin films are gaining attention due to their applicability in memory, sensor, actuator, microelectromechanical system. In this work, $Na_{0.4}K_{0.1}Bi_{0.5}TiO_3$ (NKBT0.1) ferroelectric thin films were deposited on Pt(111)/Ti/SiO2/Si substrates using the pulsed laser deposition technique at various substrate temperatures (600-750 °C). The comprehensive structural, microstructural, and ferroelectric properties characterizations depicted that the grain size, dielectric constant, and remnant polarization of the films increased with higher deposition temperatures. The influence of higher substrate temperatures on the control of [100]-preferential orientations was observed, indicating the importance of deposition conditions. Significantly, films deposited at 700 °C exhibited reduced dielectric loss of ~0.08 (at 1kHz), high dielectric constant ~673, and remnant polarization of ~17 μC/cm$^2$ at room temperature. At this deposition temperature, maximum effective piezoelectric coefficient ($d_{33}^*$) of ≈76 pm/V was availed. Based on the structural analysis, dielectric properties, and ferroelectric behavior, the optimal deposition temperature for the NKBT0.1 thin films was determined to be 700 °C. This study contributes to the understanding of the influence of substrate temperature on the structural and ferroelectric properties of Pb-free NKBT0.1 thin films, providing insights for the development of high-performance ferroelectric devices.

**# equal contribution**
Corresponding authors E-mail: [*]anbp.phy@gmail.com (A. H. Pandey)
                                [†]ramani@iitb.ac.in (N. Venkataramani)

The wide range of applications of the ferroelectric material in sensor, actuator, transducer, random access memory etc., has propelled the research and development in this material. Several lead (Pb)-containing ferroelectrics, such as $Pb(Zr_xTi_{1-x})O_3$ and $Pb(Mg_{1/3}Nb_{2/3})O_3$-based compositions have been widely studied due to their excellent ferroelectric and piezoelectric properties. Due to the toxicity of Pb, RoHS directive limits the use of Pb, which boosts up the study of the Pb-free ferroelectrics. Among the Pb-free alternatives, Bi-containing ferroelectrics for example, $Na_{0.5}Bi_{0.5}TiO_3$ and $K_{0.5}Bi_{0.5}TiO_3$ -based compositions are getting attention due to the similar kind of electronic configuration of $Bi^{3+}$ and $Pb^{2+}$, as well as high polarization value [1]. It is often found that the solid solution of $Na_{0.5}Bi_{0.5}TiO_3$ and $K_{0.5}Bi_{0.5}TiO_3$ is designed to improve the ferroelectric, piezoelectric and dielectric properties. In the bulk (1-x) $Na_{0.5}Bi_{0.5}TiO_3$-$xK_{0.5}Bi_{0.5}TiO_3$ system, x=0.2 composition (denoted as NKBT0.1) shows enhanced ferroelectric, piezoelectric and dielectric properties. This composition shows excellent piezoelectric constant ($d_{33}$) of ~215 pC/N, and remnant polarization ($P_r$) of ~ 56.7 $\mu C/cm^2$ [2]. Not only the bulk compositions, but the Pb-free piezo/ferroelectric thin films are widely studied for their applicability in the miniaturized devices. Because of the fast advancement in the miniaturization in sensor, actuators, microelectromechanical systems and other devices, it become necessary to scaling down of the piezo/ferroelectric materials [3, 4]. Due to excellent piezoelectric and ferroelectric properties, thin films of Pb-free NKBT-based ferroelectrics are vastly studied and explored for the applications in, microelectromechanical systems [5], energy storage [6,7] and pyroelectric devices [8]. It is worthy to mention here that in most of the cases sol-gel technique has been used to grow the NKBT film [9-12]. Although the sol-gel technique is commonly employed for the growth of NKBT films, this study explores the deposition of NKBT0.1 thin films using pulsed laser deposition (PLD) due to its advantages in maintaining the stoichiometry ratio [13,14]. As the significantly improved properties are obtained in the ferroelectric thin films with preferred orientation [15,16] in this work textured NKBT0.1 thin films are prepared on Pt(111)/Ti/SiO$_2$/Si substrate by keeping $La_{0.5}Sr_{0.5}CoO_3$ as a seed layer and varying the substrate temperatures. The effect of deposition temperature on the structure, microstructure, ferroelectric and dielectric properties of the NKBT0.1 thin film has been presented and discussed.

Na$_{0.4}$K$_{0.1}$Bi$_{0.5}$TiO$_3$ (NKBT0.1) target had been prepared by solid state reaction method by using precursors: Na$_2$CO$_3$ (99.9%), K$_2$CO$_3$ (99.9%), Bi$_2$O$_3$ (99.99%), and TiO$_2$ (99.9%). The details of the processing of bulk NKBT0.1 can be found in reference [17]. NKBT0.1 thin films were deposited on Pt(111)/Ti/SiO$_2$/Si substrate PLD technique using KrF eximer laser (Lambda Physik COMPex Pro) having the wavelength of 248 nm. To facilitate perovskite phase growth, a conducting buffer layer of La$_{0.5}$Sr$_{0.5}$CoO$_3$ (LSCO) thin film (~50 nm) was initially deposited on the Pt(111)/Ti/SiO2/Si substrate at 600°C, with an oxygen partial pressure of ≈ 0.133 mbar. The deposition was done with the fluence of 2 J/cm$^2$ and a repletion rate of 2 Hz. The LSCO target was kept a distance of 45 mm from the substrate. On top of LSCO, NKBT0.1 films were deposited at different substrate temperatures 600, 650, 700 and 750 °C annealed at the same temperatures for 15 minutes and then cooled to room temperature at the rate of 4 °C/min. During the deposition process the oxygen partial pressure of ≈ 0.2 mbar was kept. The fluence and repetition rate for the deposition of NKBT0.1 film were maintained 2 J/cm$^2$ and 2 Hz, respectively. The phase formation of the films deposited at different substrate temperatures were checked by by X-ray diffractometer (Rigaku) using Cu K$_\alpha$ radiation (λ = 1.5406 Å). Room temperature Raman spectra of the samples were investigated by using Raman spectrometer (Horiba). A field emission scanning electron microscope (JEOL JSM-6340F, Japan) was used to observe both the surface morphologies and cross sectional images of the films. The ferroelectric domain switching at room temperature was examined by piezoresponse force microscopy (Asylum/Oxford instruments, MFP3D origin). Au/Cr was sputtered using a shadow mask on the top of the NKBT0.1film for electrical measurements. The room temperature dielectric constant and loss were measured in alpha-A high resolution frequency analyser (Novacontrol GmbH, Germany) in the frequency range of $10^2$-$10^5$ Hz. The room temperature polarization vs electric field measurements of NKBT0.1 thin films were performed at 1 kHz using a polarization vs electric field loop measuring unit (aixACCT GmbH, TF 2000).

Fig 1(a) shows the XRD patterns of NKBT0.1 films deposited on Pt(111)/Ti/SiO$_2$/Si substrates at different substrate temperatures. The existence of secondary phase (denoted by asterisk symbol in Fig 1(a)) is noticed, when the sample is deposited at 750°C. The possible reason of the appearance of the secondary phase (Bi$_4$Ti$_3$O$_{12}$) is the volatilization of potassium at higher deposition temperature. The film deposited at 700 and 750 °C, shows the [100] preferential orientation. The texture is evidenced by the higher relative intensity of (100) Braggs reflection for the substrate temperatures of 700 and 750 °C.

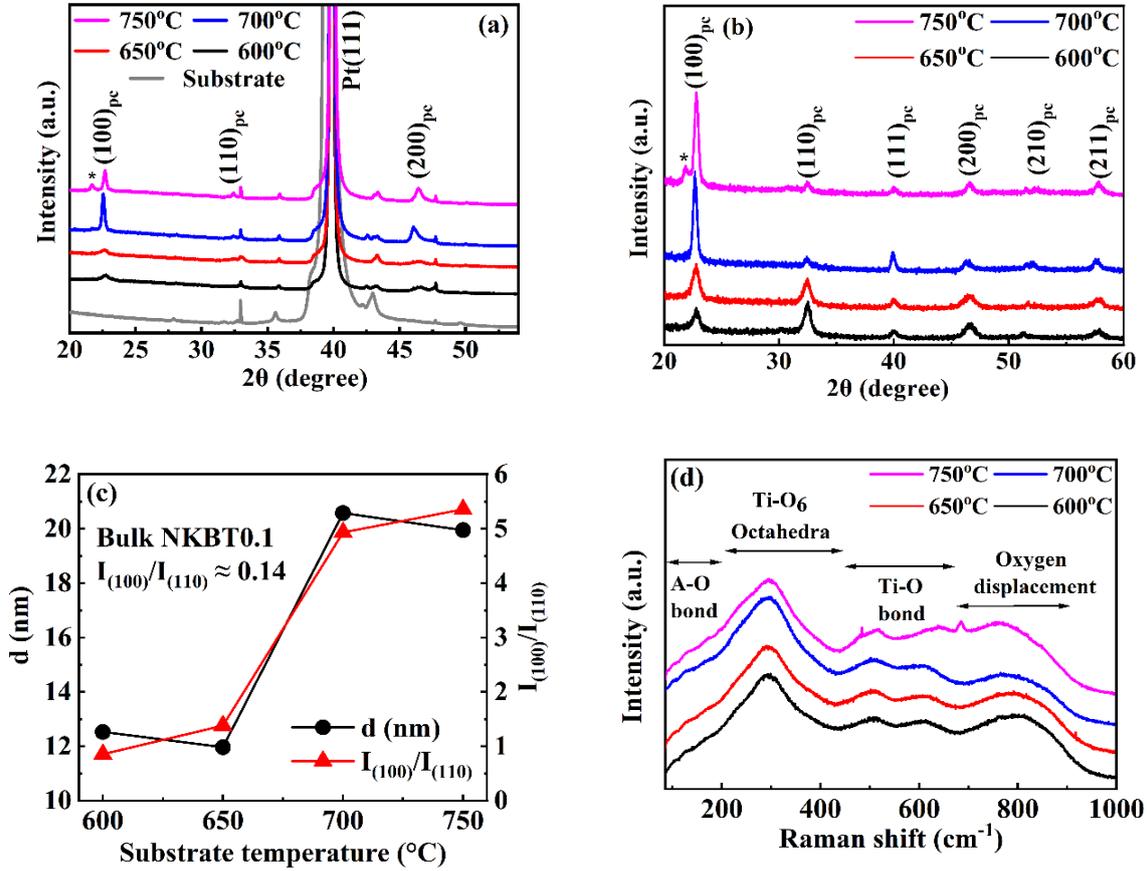

**Fig 1.** (a) XRD patterns of NKBT0.1 films deposited on Pt(111)/Ti/SiO$_2$/Si substrates at different substrate temperatures. (b) GIXRD patterns of the thin film sample. (c) The plot of intensity ratio of (100) to (110) peak and '$d$' as a function of deposition temperature. (d) Room temperature Raman spectra of NKBT0.1 THIN films.

In addition, the grazing incidence XRD (GIXRD) patterns of the NKBT0.1 thin films deposited at 700°C and 750°C also exhibit a higher relative intensity of the (100) peak (**Fig. 1(b)**). The substrate temperature also affects the crystallite size ($d$), which is evaluated by using Debye-Scherrer formula:

$$d = \frac{k\lambda}{\beta cos\theta} \tag{1}$$

where $k$ is the Scherrer constant, $\lambda$ is the wavelength (=1.541 Å) of X-rays, $\beta$ is the full width half maximum (FWHM) of (100) peak. The increment in '$d$' with the increasing deposition temperature is shown in Fig 1(c). A significant decrease in the FWHM of the (100) Bragg reflection is observed at higher substrate temperatures, indicating enhanced crystallinity. The intensity ratio of (100) to (110) peak is denoted by $I_{(100)}/I_{(110)}$ and it is plotted vs deposition temperature (Fig. 1(c)). The $I_{(100)}/I_{(110)}$ of ≈ 0.14 is calculated from the XRD data of the bulk NKBT0.1 target. However, in the case of thin film the (100) peak gradually becomes intense with the increasing deposition temperature. The highest value of $I_{(100)}/I_{(110)}$ is noticed for the film deposited at 750 °C. The [100]-preferential orientation at higher deposition temperatures can be attributed to the lowest surface energy of (100) plane [18]. As the orientation of the film parallel to the plane with minimum surface energy is always favourable, therefore minimum surface energy of the crystallographic plane influence the texture of the film [19]. At higher substrate temperature, the mobility of the adatoms on the substrate enhances, which enhances the grain growth with the lowest surface energy [20]. The larger grain size at higher substrate temperature supports this observation. Fig. 1(d) compares the room temperature Raman spectra of the samples. Presence of major bands are observed in the region of 200 to 400 $cm^{-1}$, 435 to 672 $cm^{-1}$, and 700 to 945 $cm^{-1}$. The prominent 280 $cm^{-1}$ peak in the 200 to 400 $cm^{-1}$ region is associated with the Ti-O vibration [21]. The bands in the 435 $cm^{-1}$-672 $cm^{-1}$ range are assigned to the vibration of $TiO_6$ octahedral [22], and the band situated in the 672 $cm^{-1}$ -930 $cm^{-1}$ region arise due to oxygen displacements. The surface morphologies of NKBT0.1 thin films deposited at different temperatures are shown in fig. 2(a-d). Fig. S1(a-d) of the supplementary file shows the cross section images of LSCO and NKBT0.1 films. The obtained values of the thickness of LSCO and NKBT0.1 layers are ≈ 50 nm, and 300 nm, respectively. The effect of the deposition temperature on the surface morphology is observed in this figure. As shown in Fig. 2(c) and (d), larger grains are obtained for the (100)-oriented films for higher substrate temperature of 750 °C.

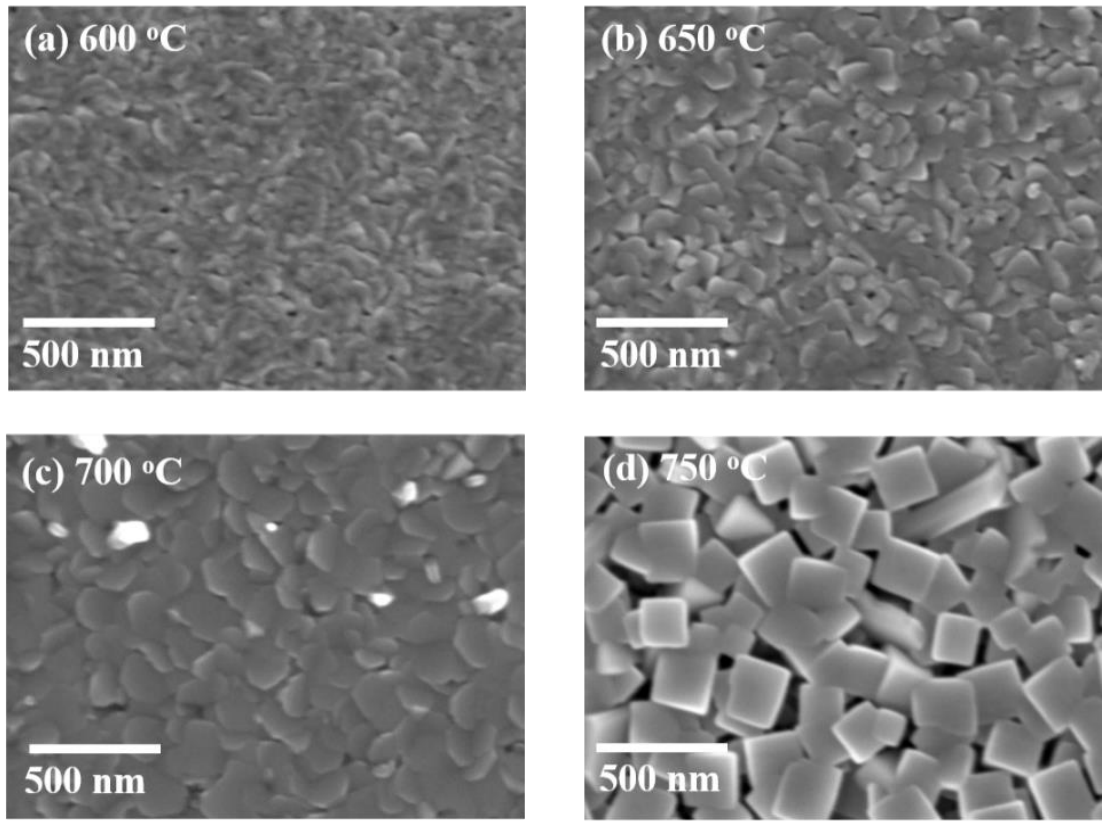

**Fig. 2.** Surface morphologies of NKBT0.1 thin films deposited at (a) 600, (b) 650, (c) 700 and (d) 750 °C.

Fig. 3(a) and (b) show the room temperature permittivity (ε′) and dielectric loss (tanδ) in the frequency (f) range of $10^2$-$10^5$ Hz of the thin films deposited at different temperatures. The higher values of ε′ are noticed for textured samples. At room temperature and 1 kHz, ε′ enhances from ≈ 451 to 700 with the increasing deposition temperature from 600 to 750 °C. The larger grain size of the film is the reason for the higher ε′ with increasing deposition temperature. The ease domain wall mobility for larger grain size enhances the ε′ [23]. Similar to ε′, tanδ also increases with the increasing deposition temperature (Fig. 3(b)). To identify the effect of substrate temperature on the ferroelectric property the polarization vs electric field (P-E) loops (Fig. 3(c)) are measured. The film deposited at 600 °C shows slimmer P-E loop, and with the increasing deposition temperature the P-E loop opens gradually. The higher remnant polarization ($P_r$) is noticed for the deposition temperature of 750 °C.

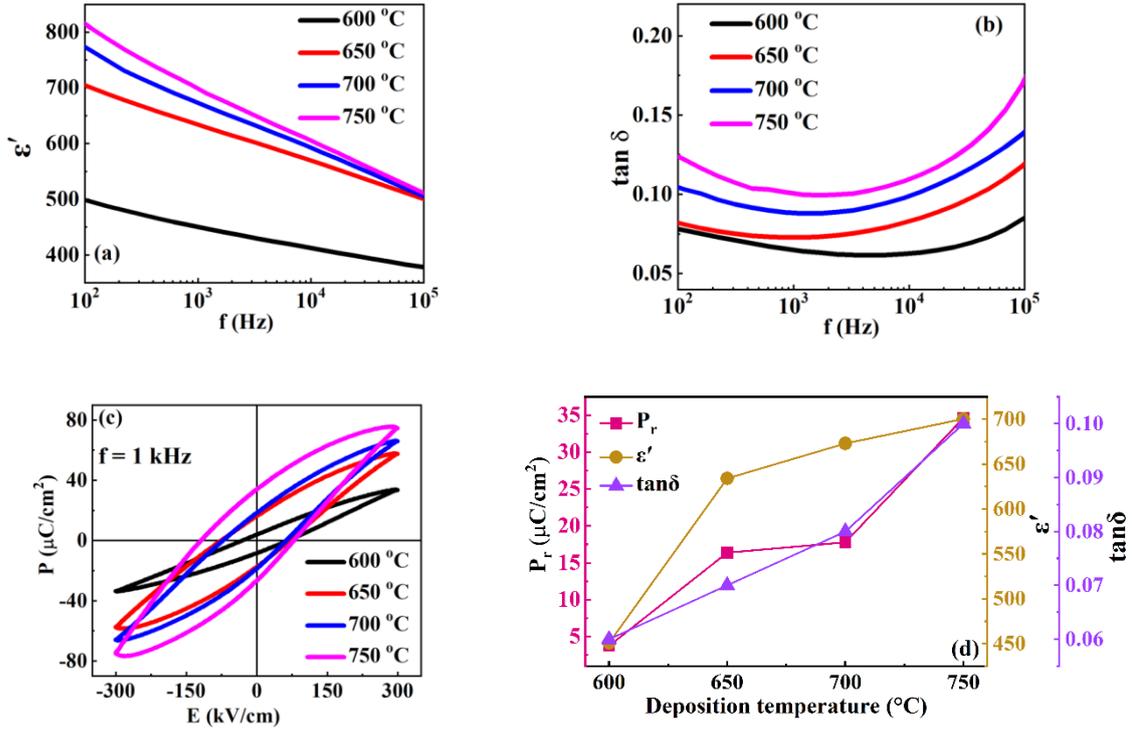

**Fig. 3:** Frequency dependent (a) ε′ and (b) dielectric loss (tanδ) of NKBT0.1 thin films deposited at different substrate temperatures. (c) Room temperature P-E loop and (d) variation of $P_r$ and ε′, tanδ values of NKBT0.1 thin films at different substrate temperatures.

Along with the [100] preferential orientation, the larger grain size is also the reason behind the obtained higher $P_r$ for higher substrate temperature [16]. In the case of larger grain, the easy switching of dipoles and reduced pinning effect of domain wall enhance the $P_r$. The variation in ε′ and tanδ at 1 kHz along with $P_r$ as a function of substrate temperature are plotted in Fig. 3(d). As shown in this figure film deposited at 750 °C has higher $P_r$, but higher dielectric loss is noticed in this deposition temperature. On the other hand, high ε′ and $P_r$ values with low dielectric loss (tanδ ≈ 0.08 at 1 kHz) are availed when the substrate temperature is 700 °C (Fig. 3(d)). Piezoresponse force microscopy (PFM) has been employed to study the polarization reversal in the samples as a function of voltage. Fig. 4(a) and (b) shows the PFM topography and phase of the NKBT0.1 film deposited at 700 °C, respectively. The PFM topography and phase of other samples are shown in Figure S2 and S3, respectively. The surface topography denotes rough surface of the NKBT0.1 films for all deposition temperatures. The bright and dark regions are observed in the PFM phase images for all samples of different substrate temperatures. The color contrast in the phase images denote the different orientation of domain in NKBT0.1.

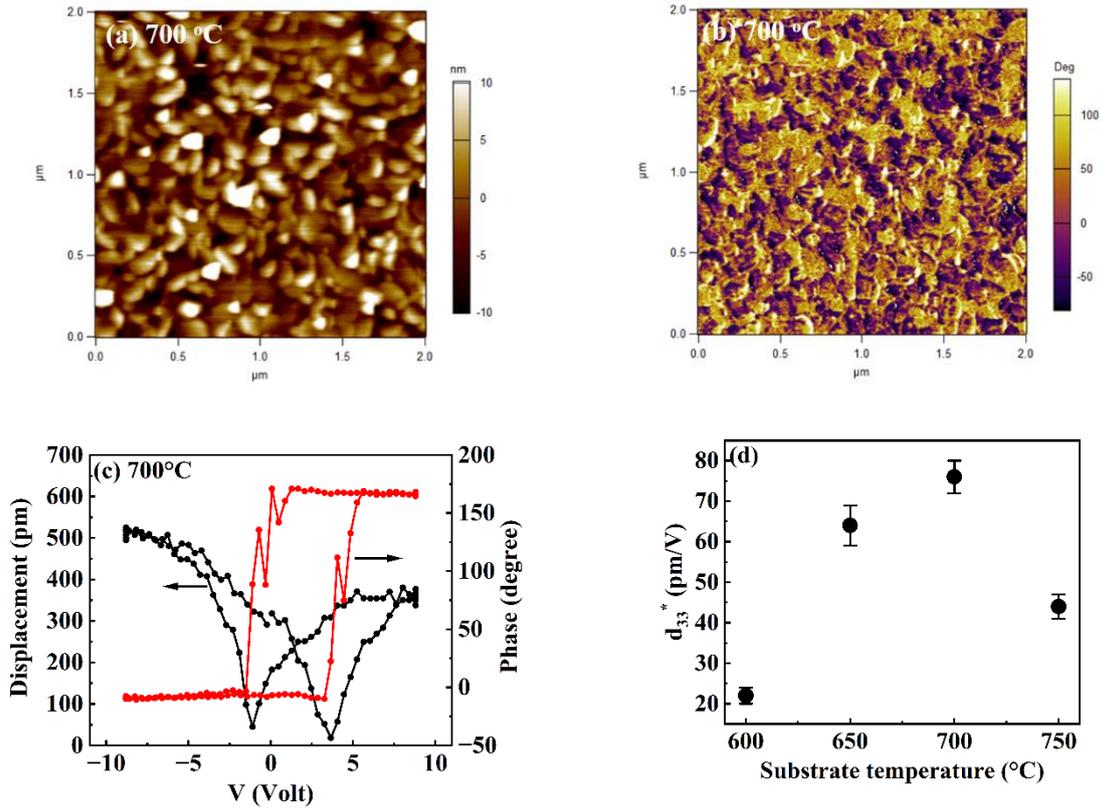

**Fig. 4:** PFM (a) topography and (b) phase contrast of the NKBT0.1 thin film deposited at 700 °C. (c) The plot of phase and displacement–voltage hysteresis loops of the same sample. (d) The graph of effective piezoelectric coefficient ($d_{33}^*$) of the films vs substrate temperature.

Fig. S4 exhibits the PFM phase image of the same sample under the presence of bias voltage. As displayed in the figure, in the outer square +20 V and in the inner square -20 V are applied. These squares show a clear color contrast, which indicates the domain switching in the sample. The domain switching is also supported by its phase vs voltage curve (Fig. 4(c)), which shows a 180° polarization reversal with a hysteresis nature. The effective piezoelectric coefficient ($d_{33}^*$) of the films are measured from the displacement vs applied voltage curve, which exhibits a butterfly loop as observed from Fig. 4(c). The displacement and phase vs applied voltage graphs of the sample deposited at 600 °C, 650 °C, and 750 °C in Fig. S5. The obtained $d_{33}^*$ values of the NKBT0.1 films are plotted as a function of substrate temperatures (Fig. 4(d)). For deposition at 700 °C, maximum $d_{33}^*$ of ≈ 76 pm/V is availed. Based on the high $d_{33}^*$, $\varepsilon'$, and $P_r$ values with low dielectric loss, it can be concluded that 700 °C is the suitable deposition temperature to obtain superior dielectric and ferroelectric properties. The obtained $d_{33}^*$, $\varepsilon'$, and $P_r$ values of this film are compared with other $(Na_{0.85}K_{0.15})_{0.5}Bi_{0.5}TiO_3$-based thin films and listed in Table 1. As shown in the table the $d_{33}^*$, $\varepsilon'$, and $P_r$ values of NKBT0.1 film deposited at 700 °C are comparable with other $(Na_{0.85}K_{0.15})_{0.5}Bi_{0.5}TiO_3$-based multilayer and B-site substituted $(Na_{0.85}K_{0.15})_{0.5}Bi_{0.5}TiO_3$ films. It denotes that applicability of NKBT0.1 textured film in the area of environmental friendly piezoelectric devices.

**Table 1:** Comparison of room temperature $d_{33}^*$, $\varepsilon'$, and $P_r$ values of NKBT0.1 films with other $(Na_{0.85}K_{0.15})_{0.5}Bi_{0.5}TiO_3$-based thin films

| Composition | Preparation method | $d_{33}^*$ (pm/V) | $\varepsilon'$ | $P_r$ ($\mu C/cm^2$) | References |
|---|---|---|---|---|---|
| $Bi_{0.5}(Na_{0.7}K_{0.2}Li_{0.1})_{0.5}]TiO_3$ | PLD | 64 | - | 13.9 | [13] |
| (100)-oriented Sc doped $(Na_{0.85}K_{0.15})_{0.5}Bi_{0.5}TiO_3$ thin film with the insertion of $LaNiO_3$ layer | Sol-gel method | 82 | 523 at 1 kHz | 26.2 | [15] |
| $(Na_{0.85}K_{0.15})_{0.5}Bi_{0.5}TiO_3$ | Aqueous sol-gel method | 76 | 519 at 100 kHz | 8.3 | [24] |
| Multilayer $(Na_{0.85}K_{0.15})_{0.5}Bi_{0.5}TiO_3$ thin film | Modified aqueous sol-gel method | 64 | 463 at 100 kHz | 18.3 | [3] |
| $(Na_{0.85}K_{0.15})_{0.5}Bi_{0.5}TiO_3$ films with $BaTiO_3$ interlayers | Aqueous sol-gel method | 75 | - | 22.1 | [4] |
| $(Na_{0.85}K_{0.15})_{0.5}Bi_{0.5}Ti_{(1-x)}Sc_xO_3$ x = 0.25 | Aqueous sol-gel method | 67 | - | 18.62 | [25] |
| $TiO_2$ coated $(Na_{0.85}K_{0.15})_{0.5}Bi_{0.5}TiO_3$ Composite film | Aqueous sol-gel method | 82(±4) | - | 24.2 (±1.2) | [16] |
| Sc-doped $(Na_{0.85}K_{0.15})_{0.5}Bi_{0.5}TiO_3$ epitaxial film | Aqueous sol-gel method | 92 | 562 at 100 kHz | 26.14 | [26] |
| NKBT0.1 (deposition temperature 700 °C) | PLD | 76(±4) | ≈ 673 at 1 kHz | 17.8 | This work |

In summary, the NKBT0.1 thin film is deposited on the Pt(111)/Ti/SiO$_2$/Si substrate and LSCO is used as a buffer layer. The results demonstrate that the substrate temperature plays a crucial role in determining the structure, microstructure, and ferroelectric properties of the NKBT0.1 thin film. The films deposited at 700 and 750 °C exhibit a [100]-preferential orientation. However, it is observed that the deposition temperature of 750 °C leads to higher dielectric loss and the presence of a secondary phase. On the other hand, the film deposited at 700 °C shows lower dielectric loss, higher $P_r$, and $\varepsilon'$ without the presence of any secondary phase. Furthermore, the deposition at 700 °C yields the maximum $d_{33}^*$ value, indicating enhanced piezoelectric performance of the film. Based on the ferroelectric and piezoelectric properties, it is concluded that the NKBT0.1 thin film deposited at 700 °C is suitable for the fabrication of Pb-free miniaturized piezoelectric devices. This deposition condition offers a favourable combination of higher Pr, low dielectric loss, and superior piezoelectric performance, making it promising for applications in miniaturized devices requiring efficient piezoelectric functionality. Further research can focus on optimizing the deposition parameters and exploring the performance of NKBT0.1 thin films in specific device applications.

**Acknowledgement**

Authors are grateful to SAIF and IRCC IIT Bombay for providing XRD, FE-SEM, Raman spectroscopy, broadband dielectric spectrometer, PE loop tracer, and PFM measurement facilities. KB and AP acknowledges Indian Institute of Bombay, Mumbai for the postdoctoral research fellowship. NV & ARK acknowledge Department of Science and Technology, India (Project Code No.: RD/0118-DST000-020) for supporting this work.

**References**

[1] F. Li, X. Hou, J. Wang, H. Zeng, B. Shen, J. Zhai, Journal of the European Ceramic Society 39 (2019) 2889–2898.

[2] M. Veera Gajendra Babu, S.M. Abdul Kader, M. Muneeswaran, N.V. Giridharan, D. Pathinettam Padiyan, B. Sundarakannan, Materials Letters 146 (2015) 81–83.

[3] Y. Wu, S.W. Or, Materials & Design 149 (2018) 153–164.

[4] Y. Wu, X. Wang, C. Zhong, L. Li, Journal of the European Ceramic Society 38 (2018) 1434–1441.

[5] L. Xujun, P. Yong, G. Yueqiu, H. Renjie, L. Jiajia, X. Shuhong, Z. Yichun, G. Xingsen, J Mater Sci: Mater Electron 25 (2014) 1416–1422.

[6] Y. Han, J. Qian, C. Yang, Ceramics International 45 (2019) 22737–22743.

[7] J. Wang, Y. Li, N. Sun, Q. Zhang, L. Zhang, X. Hao, X. Chou, Journal of Alloys and Compounds 727 (2017) 596–602.

[8] Q. Chi, J. Dong, C. Zhang, X. Wang, Q. Lei, J. Mater. Chem. C 4 (2016) 4442–4450.

[9] Q.G. Chi, J.F. Dong, C.H. Zhang, Y. Chen, X. Wang, Q.Q. Lei, Journal of Alloys and Compounds 704 (2017) 336–342.

[10] T. Zhang, W. Zhang, Y. Chen, J. Yin, Optics Communications 281 (2008) 439–443.

[11] Y. Wu, X. Wang, C. Zhong, L. Li, Thin Solid Films 519 (2011) 4798–4803.

[12] Q.G. Chi, F.Y. Yang, C.H. Zhang, C.T. Chen, H.F. Zhu, X. Wang, Q.Q. Lei, Ceramics International 39 (2013) 9273–9276.

[13] D.Y. Wang, D.M. Lin, K.S. Wong, K.W. Kwok, J.Y. Dai, H.L.W. Chan, Applied Physics Letters 92 (2008) 222909.

[14] A.H. Pandey, K. Miriyala, P. Varade, N.S. Sowmya, A.R. Kulkarni, N. Venkataramani, AIP Conference Proceedings 2265 (2020).

[15] Y. Wu, Y. Hu, X. Wang, C. Zhong, L. Li, RSC Adv. 7 (2017) 44136–44143.

[16] Y. Wu, X. Wang, W. Cui, Y. Hu, L. Li, Journal of the European Ceramic Society 39 (2019) 269–276.

[17] A.H. Pandey, P. Varade, K. Miriyala, N.S. Sowmya, A. Kumar, A. Arockiarajan, A.R. Kulkarni, N. Venkataramani, Materials Today Communications 26 (2021) 101898.

[18] W. Gong, J.-F. Li, X. Chu, L. Li, Journal of the American Ceramic Society 87 (2004) 1031–1034.

[19] C.V. Thompson, Annu. Rev. Mater. Sci. 30 (2000) 159–190.

[20] K.-F. Chiu, F.C. Hsu, G.S. Chen, M.K. Wu, J. Electrochem. Soc. 150 (2003) A503.

[21] J. Kreisel, A.M. Glazer, G. Jones, P.A. Thomas, L. Abello, G. Lucazeau, Journal of Physics Condensed Matter 12 (2000) 3267–3280.

[22] C. Chen, L. Yang, X. Jiang, X. Huang, X. Gao, N. Tu, K. Shu, X. Jiang, S. Zhang, H. Luo, Crystals 10 (2020) 435.

[23] C.A. Randall, N. Kim, J.-P. Kucera, W. Cao, T.R. Shrout, Journal of the American Ceramic Society 81 (2005) 677–688.

[24] W. Cui, X. Wang, L. Li, Ceramics International 41 (2015) S37–S40.

[25] Y. Wu, X. Wang, L. Li, Journal of the American Ceramic Society 94 (2011) 2518–2522.

[26] Y.Y. Wu, X.H. Wang, C.F. Zhong, L.T. Li, Key Engineering Materials 697 (2016) 235–238.

# Supplementary file

**Substrate temperature-dependent dielectric and ferroelectric properties of (100) – oriented lead-free $Na_{0.4}K_{0.1}Bi_{0.5}TiO_3$ thin films grown by pulsed laser deposition**

Krishnarjun Banerjee[#], Adityanarayan H. Pandey[‡#], Pravin Varade, Ajit R. Kulkarni, Abhijeet L. Sangle, N. Venkataramani[§]

Department of Metallurgical engineering and Materials Science, Indian Institute of Technology Bombay, Mumbai-400076, India.

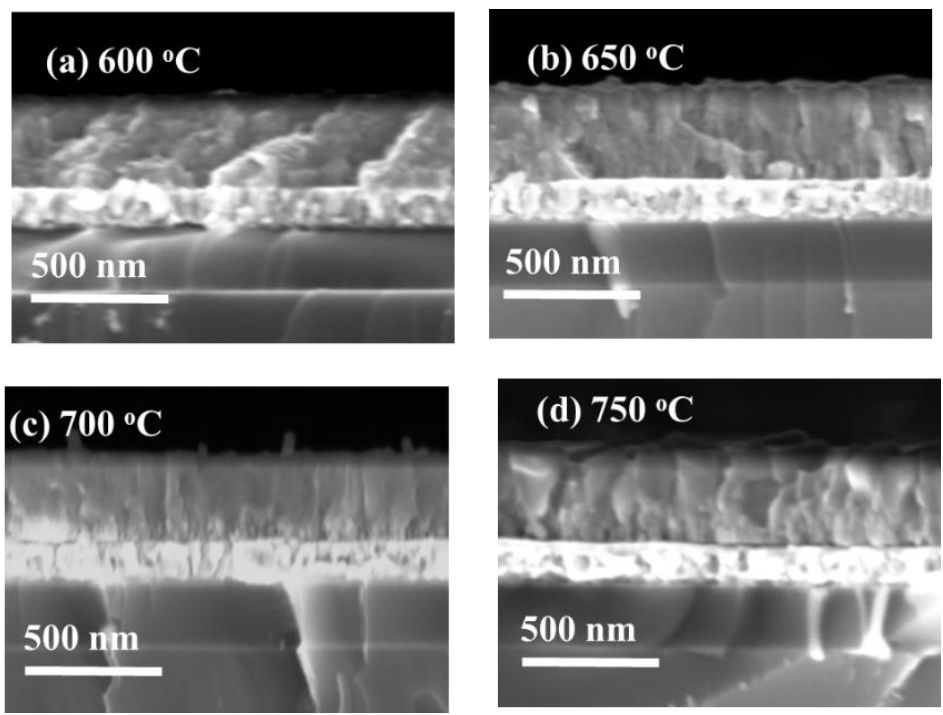

**Fig. S1(a-d)** The cross section images of LSCO and NKBT0.1 thin films. The deposition temperatures of NKBT0.1 are mentioned in the figures.

# equal contribution
Corresponding authors E-mail: [‡]anbp.phy@gmail.com (A. H. Pandey)
[§]ramani@iitb.ac.in (N. Venkataramani)

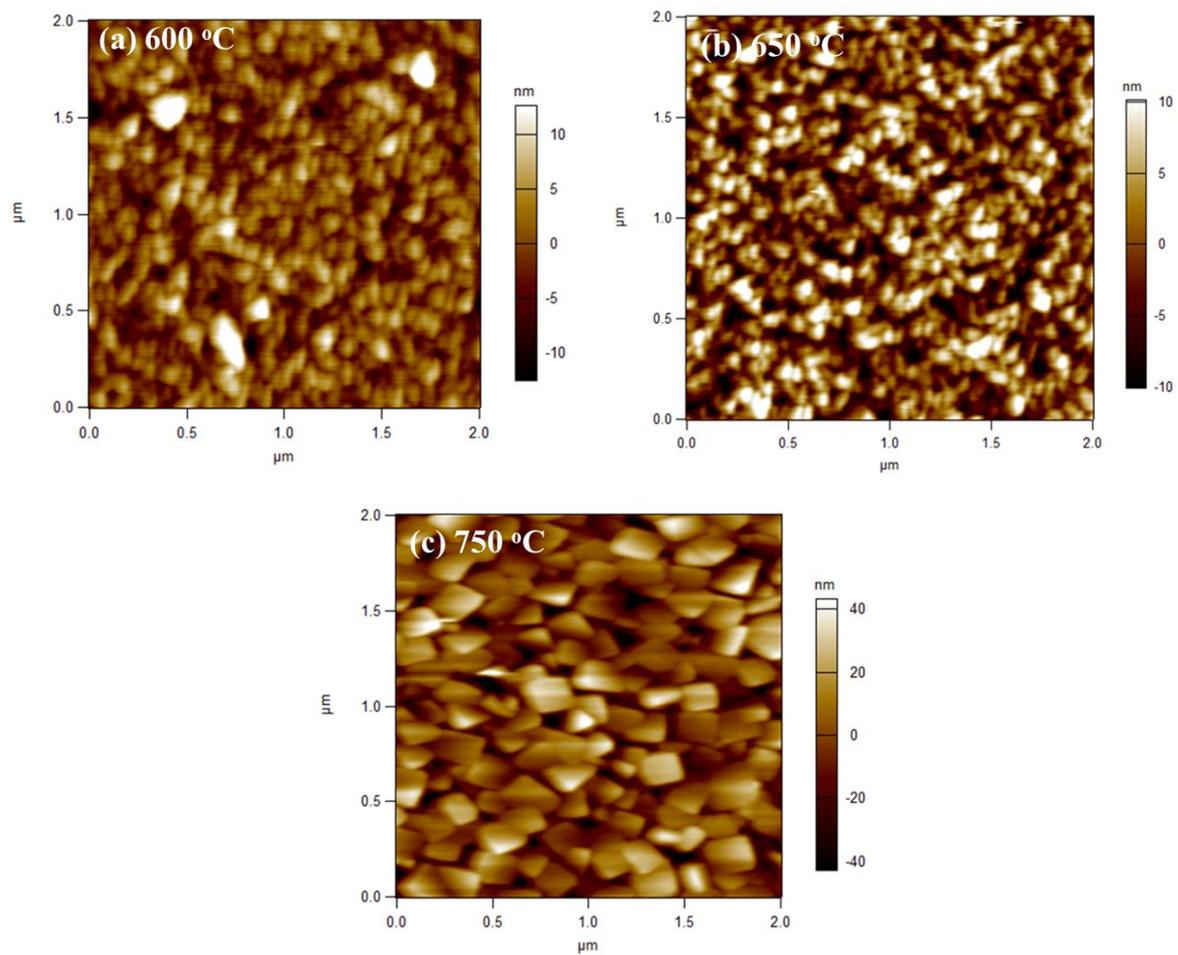

**Fig S2.** The PFM topography of NKBT0.1 thin films deposited at (a) 600, (b) 650, and (c) 750 °C.

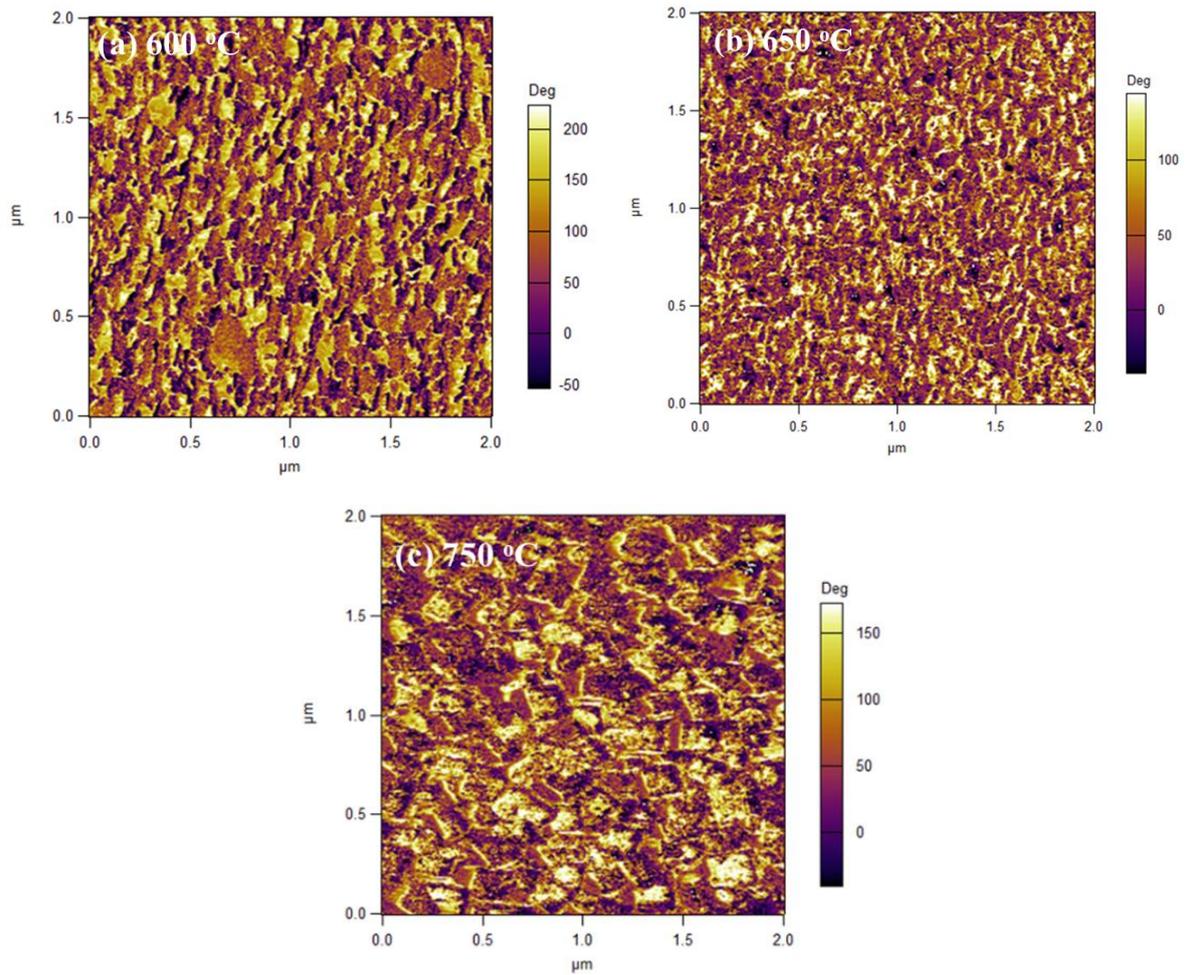

**Fig S3.** The PFM Phase image of NKBT0.1 thin films deposited at (a) 600, (b) 650, and (c) 750 °C.

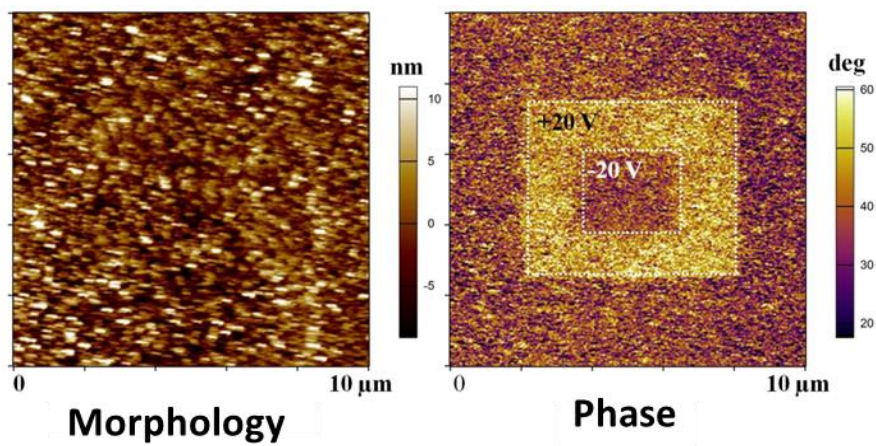

**Fig S4.** PFM morphology and phase image of NKBT0.1 film deposited at 700 °C by applying ±20 V.

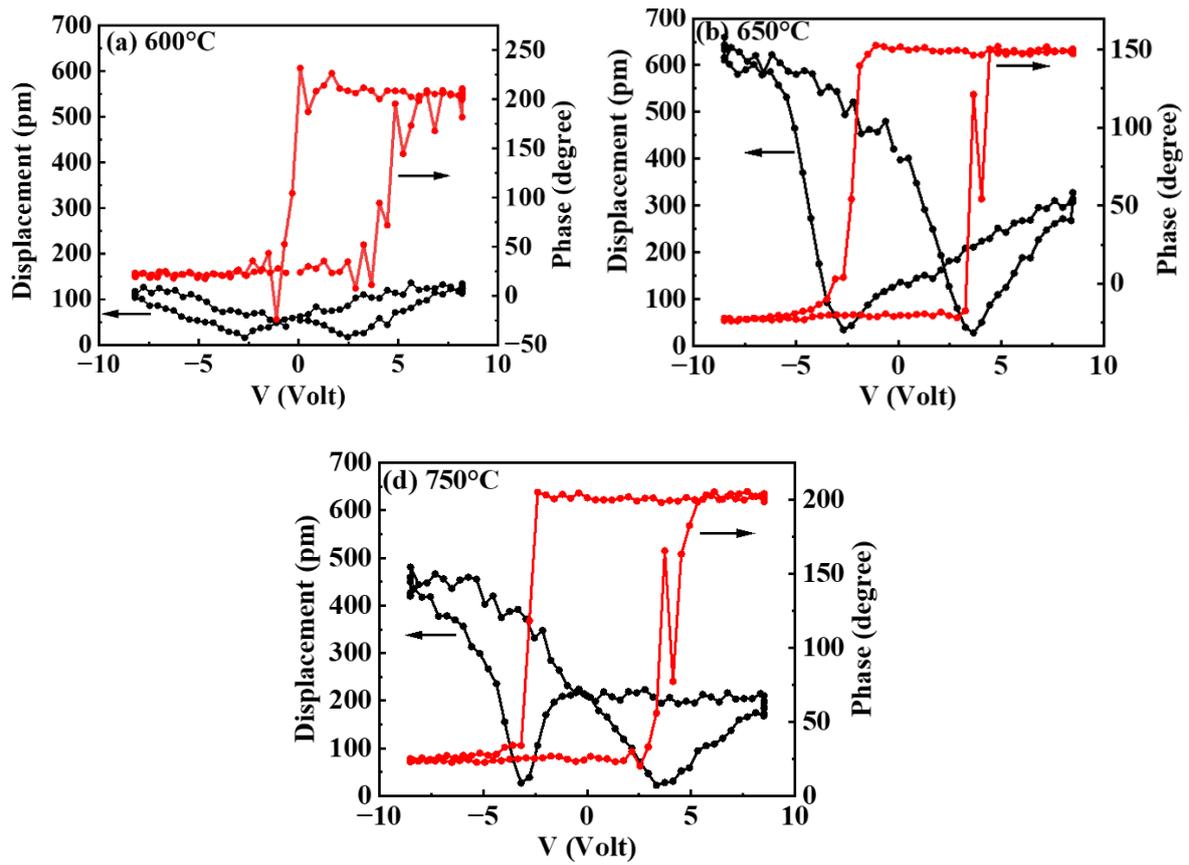

**Fig. S5.** The displacement and phase vs applied voltage hysteresis loops of thin films deposited at 600 °C, 650 °C, and 750 °C.